\documentclass[11pt,twoside]{article}
\usepackage{asp2010}
\resetcounters

\begin{document}

\title{A New Method for Measuring the Upper End of the IMF}
\author{Daniela Calzetti$^1$, Rupali Chandar$^2$, Janice C. Lee$^3$, Bruce G. Elmegreen$^4$, Robert C. 
Kennicutt$^5$, Brad C. Whitmore$^6$
\affil{$^1$Dept. of Astronomy, University of Massachusetts -- Amherst, 710 North Pleasant Street, LGRT-B 524, MA, U.S.A.; 413-545-3556; calzetti@astro.umass.edu}
\affil{$^2$Dept. of Physics and  Astronomy, University of Toledo, Toledo, Ohio}
\affil{$^3$Carnegie Observatories of Washington, Pasadena, California}
\affil{$^4$IBM T.J. Watson Research Center, Yorktown Heights, New York}
\affil{$^5$Institute of Astronomy, Cambridge University, Cambridge, U.K.}
\affil{$^6$Space Telescope Science Institute, Baltimore, Maryland}
}

\begin{abstract}
A method is presented here for investigating variations in the upper end of the stellar Initial Mass Function (IMF) by probing the production rate of ionizing photons in unresolved, compact star clusters with ages$<$10~Myr and covering a range of masses. We test this method on the young cluster population in the nearby galaxy M51a, for which multi--wavelength observations from the Hubble Space Telescope are  available. Our results indicate that the proposed method can probe the upper end of the IMF in galaxies located out to at least ~10 Mpc, i.e., a factor ~200 further away than possible by counting individual stars in young compact clusters. Our results for this galaxy show no obvious dependence of the upper mass end of the IMF on the mass of the star cluster, down to $\sim$1000 M$_{\odot}$, although more extensive analyses involving lower mass clusters and other galaxies are needed to confirm this conclusion. 
\end{abstract}

\section{Introduction}
Recent analyses of optical and UV data of nearby galaxies have clearly indicated that, as some global galactic parameter decreases (luminosity, SFR, SFR/area, etc.), so does the galaxy--integrated ionizing photon rate, as traced by the hydrogen recombination line H$\alpha$, per unit UV or optical luminosity \citep{Hove2008,Meur2009,Lee2009}. Presentations from these authors, containing also updates over the cited papers, are included in these Proceedings, together with similar results obtained by other investigators. 

In all cases, corrections for dust attenuation have been included and/or discussed, and uncertainties deriving from such corrections are unlikely to produce the observed trend over the full luminosity range analyzed \citep[factor $\approx$10$^4$ in B--band luminosity, see][]{Lee2009}. At least three possible, non--mutually--exclusive, scenarios have been proposed to account for the observed trend: 
\begin{enumerate}
\item it is an effect of stochastic sampling and/or a particular period in the star formation history of the ionizing~photon~rate--deficient (N$_{ion}$--deficient) galaxies. Specifically, these galaxies could be in a post--burst phase, and the apparent higher frequency of such phase in low--luminosity systems could be due to the sporadic nature of star formation in these systems;
\item it is an effect of ionizing photon leakage from the star--forming regions of  these low--mass and low--density systems;
\item it is an effect of a systematic behavior of the high--end of the stellar IMF linked to the variation of some global galactic parameter.
\end{enumerate}

The first scenario has been discussed by all authors cited so far, and considered unlikely to be the only or primary driver of the observed trend, given that it would require synchronization among the star formation histories of the low--luminosity galaxies (see, also, the presentation by E. Hoversten in these Proceedings). These results notwithstanding, it may still be worth investigating the potential effects of sporadic star formation with increasingly sophisticated prescriptions for the star formation histories of galaxies. Accurate star formation histories  are only now becoming available for a significant number of low--mass and low--luminosity galaxies \citep[e.g.,][]{Weis2008}. Furthermore, low--luminosity galaxies show over an order of magnitude larger variation in their birth--rate parameter than higher luminosity galaxies \citep{Lee2007}, which may possibly impact observables. 

The second scenario is discussed in, e.g., \citet{Hunt2010}. Although a potentially attractive interpretation, leakage of photons from galaxies is difficult to measure. A few observations performed in higher density systems than those under consideration here place an upper limit of a few percent to the leakage of ionizing photons from low--redshift galaxies \citep{Heck2001,Deha2001,Haye2007,Blan1999}. Similar limits of about 6\%, and no higher than 20\%, have been recently obtained for Lyman Break Galaxies at redshift $\approx$4 \citep{Vanz2010}. Ionizing photons leakage from  low--luminosity, low--mass galaxies is clearly important to establish, as it could have major implications for the re--ionization of the Universe at high--redshift. However, at low density, leakage of ionizing photons throughout a galaxy disk or into a diffuse halo would be very difficult to detect at the resulting low emission measures, even if the photons did not leave the galaxy entirely. 

The third scenario has been suggested by \citet{Pfla2009}, and builds on the model proposed by \citet{Weid2004}, \citet{Weid2005}, \citet{Weid2006}, and \citet{Weid2010}, according to which: (1) there is a correlation between a galaxy's SFR and the maximum mass of the cluster that can form in that galaxy \citep{Lars2002}; (2) there is a correlation between a cluster mass and the most massive {\em possible} star the cluster can form \citep{Weid2006}. The combination of these two conditions produces an integrated galaxial stellar IMG (IGIMF) that reproduces the observed decrease in the ionizing photon rate  per unit UV luminosity as a function of decreasing galactic SFR, if the assumption that most of the stars form in stellar clusters holds. 

\subsection{Goal of This Presentation}
This presentation, based on the results in \citet{Calz2010}, is concerned with the second of the IGIMF's requirements, i.e., that a correlation might be present between a cluster mass and the most massive star it can form. This is another way to say that the formation of stars in a cluster is not an event solely driven by stochastic sampling of the IMF \citep{Elme2006}, and 100 clusters each with mass=10$^2$~M$_{\odot}$ are not equivalent to one 10$^4$~M$_{\odot}$ cluster, as the former will systematically lack the  high--mass stars contained in the latter.

Measuring the number of massive stars in a stellar cluster is a problem that increases in difficulty with increasing distance. Significant blending will occur for massive stars due to crowding in the centers of star clusters \citep{Apel2008,Asce2009}, even with the angular resolution of the Hubble Space Telescope, for distances larger than those of the Magellanic Clouds ($\sim$55~kpc).  All the N$_{ion}$--deficient galaxies are farther away than the Magellanic Clouds. Furthermore, the labor--intensive approach of counting individual stars in single--age, young stellar populations (typically  $\lesssim$3~Myr~old star clusters), which is the standard approach for deriving stellar IMFs, cannot build statistically large samples of such measurements. 

The approach to constraining the upper end of the IMF presented here exploits the integrated properties of star clusters. 
Each star cluster is treated as a single unit, and many such units are averaged together in order to investigate variations on the upper--IMF; this is accomplished by measuring the dependence of the  ionizing photon rate, normalized to the cluster mass, 
on cluster mass. This requires that accurate ages and masses be determined for each cluster. This method 
extends the \citet{Corb2009}  approach (the cluster birthline, see also Corbelli's contribution to these Proceedings) 
by normalizing the ionizing photon rate to the age--independent cluster mass, instead of the age--dependent bolometric luminosity. 

\section{Framework}
\begin{figure}
\plottwo{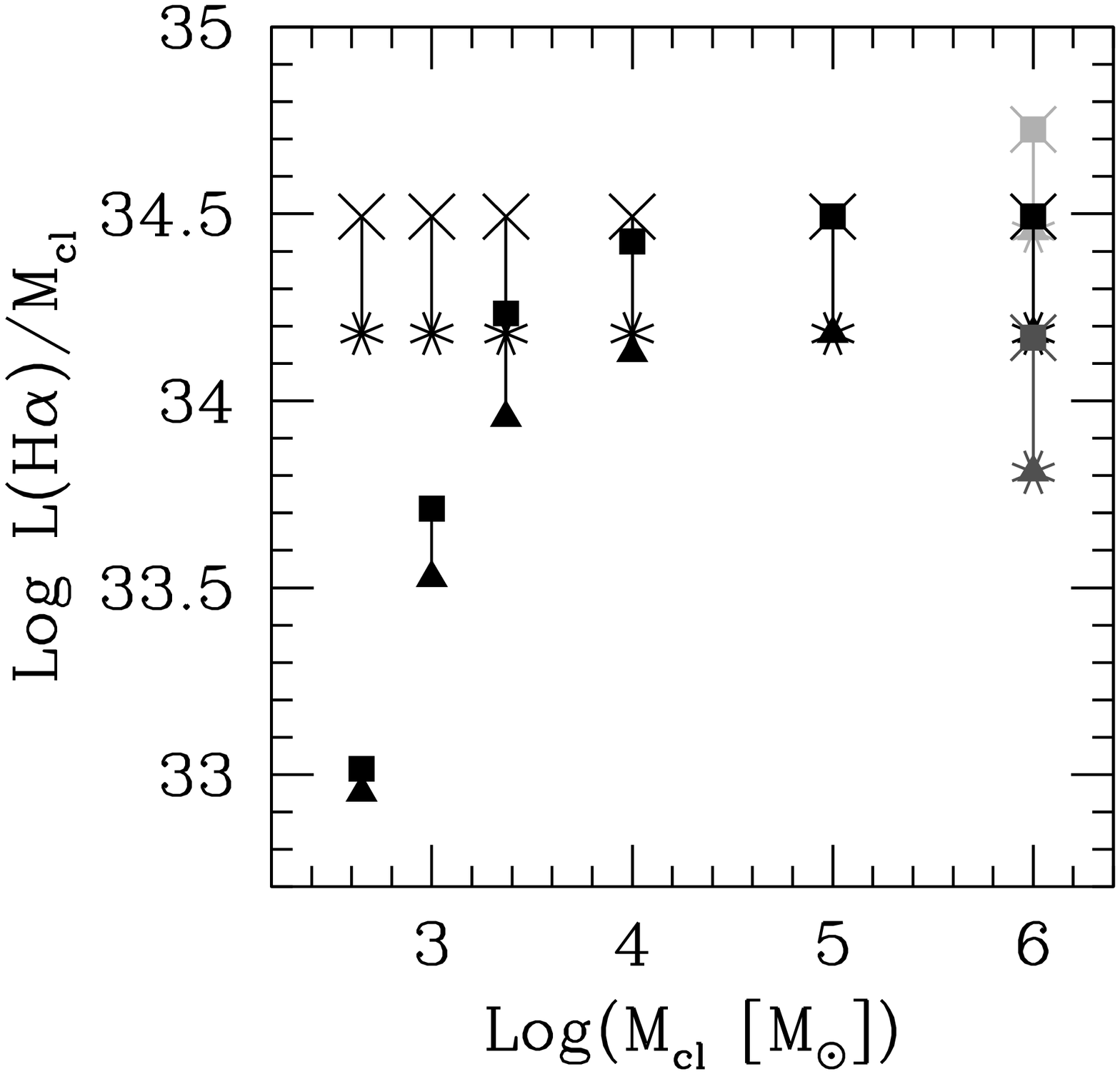}{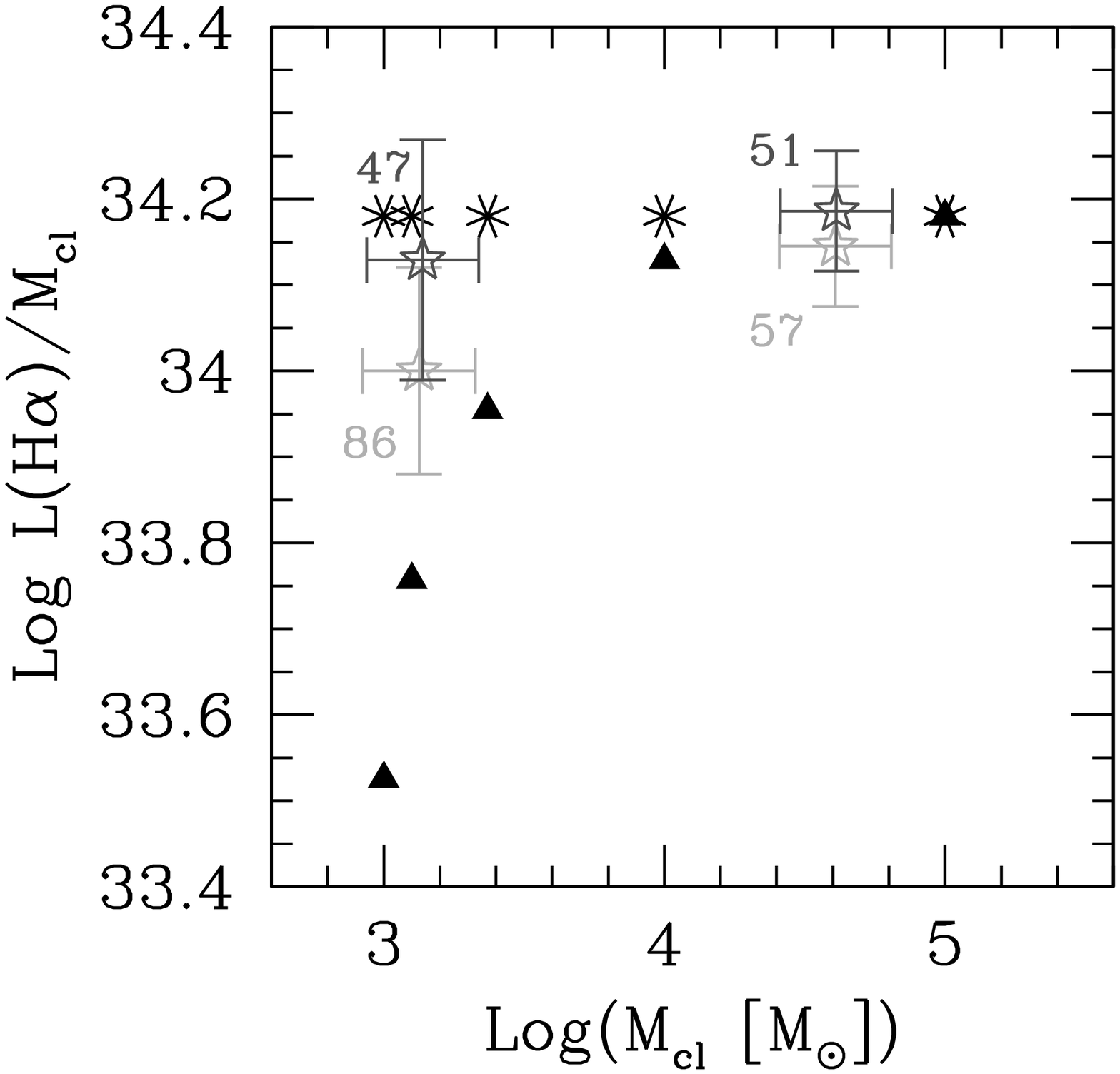}
\caption{{\bf LEFT:} The expected ratio L(H$\alpha$)/M$_{cl}$ as a function of the cluster mass M$_{cl}$ both 
for a universal IMF (crosses and asterisks), and for a cluster--mass--dependent IMF \citep[filled squares and triangles,][]{Weid2006}. The models are from Starburst99 \citep[2007 Update,][]{Leitherer1999}, and assume averages over large numbers of clusters. The mean L(H$\alpha$)/M$_{cl}$ ratios are averaged over two age ranges: 1.5--5~Myr (crosses and squares) and 2--8~Myr (asterisks and triangles), joined by vertical bars, under the assumption of constant star formation for each age range. We assume that clusters younger than 1.5--2~Myr are too dust--enshrouded to provide significant contribution at optical wavelengths. The black symbols are for 1.5~solar metallicity models; examples of higher ($\sim$3.5 solar, dark~grey) and lower ($\sim$1/3 solar, light~grey) metallicity are shown at a representative mass. 
 {\bf RIGHT:} The ratio $<$L(H$\alpha$)$>$/$<$M$_{cl}>$ as a function of the mean cluster mass M$_{cl}$ in two mass bins
  for M51a, for clusters younger than $\sim$8--10~Myr (stars with 1~$\sigma$ error bars), including all clusters within each mass bin (light~grey, bottom symbols), and including only clusters with H$\alpha$ flux detections $>$3~$\sigma$ (dark~grey, top symbols). The number of clusters in each  mass  bin is shown close to the symbols.  Expectations from models with 1.5 solar metallicity and averaged over the 2--8~Myr age range are shown (asterisks and triangles), as they are the closest match to the M51a high--mass data. 
\label{fig1}
}
\end{figure}

Measuring the ionizing photon rate, N$_{ion}$, in a young, massive star cluster is equivalent to measuring the number of massive stars that cluster contains at that age. The younger the age, the closer the measured number is to the total ionization rate of massive stars at birth (including the most massive stars formed), or the integral of the upper--IMF over mass. Star clusters also represent ideal sites for such measurements, because massive star populations tend to be coeval or close to coeval. If a trend is predicted for the number of massive stars or the most massive star formed as a function of the cluster's mass, then the presence/absence of such trend can be established by investigating the behavior of N$_{ion}$/M$_{cl}$ as a function of the cluster's mass M$_{cl}$. The left panel of Figure~\ref{fig1} shows the expected trends for a `universal' IMF, where the most massive star formed is independent of the cluster's mass, and for an IMF where the most massive star formed depends on the cluster's mass \citep[the model of][]{Weid2006}; the ionizing photon rate is here replaced with the extinction--corrected H$\alpha$ luminosity. The Figure also shows the expected range for different values of the cluster's metallicity and for different ranges of the clusters' ages.

This simple framework is fraught with a number of uncertainties, all of which need to be carefully evaluated. Even in the event that the IMF is `universal', stochastic sampling of the IMF is one of such uncertainties, which starts to be important as the mass of the cluster decreases. Below masses$\approx$a~few$\times$10$^4$~M$_{\odot}$, the IMF cannot be fully sampled, and since fractional stars cannot be created, the stars formed in the cluster will be randomly sampled from the IMF. This induces a scatter on the measured parameters. For instance, the scatter on the mass determination increases from  6\% for a 10$^5$~M$_{\odot}$ cluster to $\sim$50\% for a 10$^3$~M$_{\odot}$ cluster; the scatter on  N$_{ion}$ increases from 
$\sim$10\% to 80\% for the same cluster mass range \citep{Cerv2002,Cerv2004}. These uncertainties can be reduced by combining large numbers of same--mass, young clusters to a cumulative equivalent mass $\approx$10$^5$~M$_{\odot}$, since a `universal' IMF would be fully sampled in a cluster of this mass, and measuring the summed values of the combined clusters as L(H$\alpha$)/M$_{cl}$=$<$ L(H$\alpha$)$>$/$<$M$_{cl}$$>$. 

Stochastic sampling, when combined with photometric and dust reddening uncertainties, can also affect age determinations from multi--band photometry, by `polluting' samples of young clusters with the addition of old members or loss of young ones, and lead to mass underestimates by up to a factor $\sim$2 \citep{Chan2010b,Foue2010}. Even for massive clusters, i.e., in the absence of stochastic sampling, observational uncertainties still affect final age determinations at the level of a factor $\sim$2 and masses at the 60\% level. Even achieving this level of `accuracy' requires a full set of U,B,V,R,I photometry \citep{Chan2010a,Chan2010b}. The uncertainties in the ages can be `mimicked' by models, and be incorporated in the expectations, as shown in the left panel of Figure~\ref{fig1}, where the ratio L(H$\alpha$)/M$_{cl}$ is shown averaged over two age bins: 1.5--5~Myr and 2--8~Myr. Conversely, uncertainties in the masses are carried though, in our treatment, as observational uncertainties. 

\section{Results for the Young Cluster Population of M51a}
The diagnostic plot discussed in the previous section should be applied to the low--luminosity, low--SFR galaxies where the deficiency in ionizing photon rate has been detected, in order to test whether such deficiency may be caused by the 
`maximum star mass -- cluster mass' correlation hypothesis discussed in the Introduction. However, reliable identification and, especially, photometric measurements of the low--mass star clusters expected to be present in such small galaxies become incresingly difficult beyond $\approx$1~Mpc, with ground--based facilties (most N$_{ion}$--deficient galaxies are located beyond 1~Mpc). In general, multi--wavelength, at least U--to--I, photometry from the HST is required, but such observations are not currently available for those galaxies.

Age and mass measurements of clusters are now becoming available for more massive galaxies, where significant populations of star clusters are present and are being investigated to understand the formation and evolution of these systems \citep[e.g.,][]{Bast2005,Whit2010,Chan2010a,Chan2010b}. These populations can be used a test--beds for the validation of the method described in section~2. We chose here the cluster population of the well--studied nearby galaxy M51a (8.4~Mpc), for which multi--band HST images are available; this galaxy has super--solar metallicity \citep{Mous2010}, an important piece of information, since the ratio L(H$\alpha$)/M$_{cl}$ depends on metallicity (Figure~\ref{fig1}, left panel).

The right panel of Figure~\ref{fig1}, shows the location of the young, $\lesssim$8--10~Myr, M51a star clusters in the L(H$\alpha$)/M$_{cl}$--versus--M$_{cl}$ plane, where the H$\alpha$ luminosity has been corrected for dust attenuation, and each mass point is the sum of the number of clusters indicated in the figure. The young clusters are divided into two mass bins (star symbols), 
and the symbols represent: (a) sums made with all clusters whose colors indicate young ages  (light~grey symbols), and (b) 
sums made with only clusters which have colors of young populations {\em and}  also have H$\alpha$ detections above 3~$\sigma$ (dark~grey symbols). 

The data show that in the case of the cluster population of M51a there is no obvious differential trend in L(H$\alpha$)/M$_{cl}$ for the two mass bins; they are consistent with a universal, stochastically--sampled,  IMF, although the `maximum star mass -- cluster mass' correlation cannot be completely excluded at this stage (at the $\approx$2~$\sigma$ level). This is not the most stringent test of the stochastic sampling model because clusters more massive than 10$^3$~M$_{\odot}$ are expected to contain more than 1 O--type star.

The data follow the 2--8~Myr age average points (asterisks) of 
the 1.5~solar~metallicity stellar population models, implying the following, non--exclusive, possibilities: (1) our criteria select clusters fairly uniformly distributed in the age range 2--8~Myr; (2) the models used in Figure~\ref{fig1}, right, have too low metallicity for M51a, and a higher metallicity model should be used (cf. Figure~\ref{fig1}, left panel); (3) significant, mass--independent, ionizing photon leakage is present in our regions, which we have not attempted to correct for in our analysis. 

\section{Conclusions}
While direct measurements of  the stellar IMF, even at the upper end where stars are bright, become a daunting and ultimately 
impossible task beyond distances of tens or hundreds of kpc from our Galaxy, integral properties of massive stars in coeval populations (young star clusters) can be exploited to at least place limits on variations of the upper--IMF up to distances of at least 10~Mpc. Specifically, a diagnostic plotting the ionizing photon rate per unit star cluster mass  as a function of the cluster mass itself can provide sufficient discriminating power for some models of IMF variation. 

This method is made more attractive by the increasing availability of well--characterized cluster populations in galaxies, thanks to the angular resolution and multi--band coverage of the HST, although there is still dearth of information for the galaxies where a deficiency of ionizing photon rate has been observed.

A number of uncertainties in the age and mass determinations are still present, however, from degeneracies in the multi--band colors (which require at least coverage in 5 bands from U to I, and possibly UV) and from the stochastic sampling of the IMF for cluster masses below $\approx$10$^4$~M$_{\odot}$. These will need to be carefully quantified. 

Despite all limitations, some conclusions can be drawn at least for the young cluster population in the nearby galaxy M51a. The HST data show that the ionizing photon rate from these clusters remains constant (when divided by cluster mass) for decreasing cluster mass. This suggests that the clusters in M51a are consistent with a universal, stochastically--sampled,  IMF. This is still a $\approx$2~$\sigma$ result, and analyses of additional galaxies covering a large range of parameters and  cluster masses lower than $\sim$10$^3$~M$_{\odot}$ will be needed to place this result on firmer ground.



\end{document}